\documentclass[english,prl, reprint, superscriptaddress, secnumarabic, showkeys, showpacs, raggedbottom]{revtex4-1}
\usepackage[T1]{fontenc}
\usepackage[latin9]{inputenc}
\usepackage{amsmath}
\usepackage{amssymb}
\usepackage{graphicx}
\usepackage{esint}

\makeatletter

\providecommand{\tabularnewline}{\\}

\@ifundefined{textcolor}{}
{%
 \definecolor{BLACK}{gray}{0}
 \definecolor{WHITE}{gray}{1}
 \definecolor{RED}{rgb}{1,0,0}
 \definecolor{GREEN}{rgb}{0,1,0}
 \definecolor{BLUE}{rgb}{0,0,1}
 \definecolor{CYAN}{cmyk}{1,0,0,0}
 \definecolor{MAGENTA}{cmyk}{0,1,0,0}
 \definecolor{YELLOW}{cmyk}{0,0,1,0}
}

\usepackage[caption=false]{subfig}

\@ifundefined{showcaptionsetup}{}{%
 \PassOptionsToPackage{caption=false}{subfig}}
\usepackage{subfig}
\makeatother

\usepackage{babel}
\begin{document}
\global\long\def\s{\sigma}
\global\long\def\k{\kappa}
\global\long\def\ph{\hat{n}}
\global\long\def\aa{\hat{a}}
\global\long\def\bra#1{\left\langle #1\right|}
 \global\long\def\ket#1{\left|#1\right\rangle }
\DeclareRobustCommand{\one}{\raisebox{.5pt}{\textcircled{\raisebox{-.9pt}{1}}}}\DeclareRobustCommand{\two}{\raisebox{.5pt}{\textcircled{\raisebox{-.9pt}{2}}}}

\title{Universal Control of an Oscillator with Dispersive Coupling to a
Qubit}

\author{Stefan Krastanov}

\affiliation{Departments of Applied Physics and Physics, Yale University, New
Haven, Connecticut 06520, USA}

\author{Victor V. Albert}

\affiliation{Departments of Applied Physics and Physics, Yale University, New
Haven, Connecticut 06520, USA}

\author{Chao Shen}

\affiliation{Departments of Applied Physics and Physics, Yale University, New
Haven, Connecticut 06520, USA}

\author{Chang-Ling Zou}

\affiliation{Departments of Applied Physics and Physics, Yale University, New
Haven, Connecticut 06520, USA}

\affiliation{Key Lab of Quantum Information, University of Science and Technology
of China, Hefei 230026, Anhui, China}

\author{Reinier~W.~Heeres}

\affiliation{Departments of Applied Physics and Physics, Yale University, New
Haven, Connecticut 06520, USA}

\author{Brian Vlastakis}

\affiliation{Departments of Applied Physics and Physics, Yale University, New
Haven, Connecticut 06520, USA}

\author{Robert J. Schoelkopf}

\affiliation{Departments of Applied Physics and Physics, Yale University, New
Haven, Connecticut 06520, USA}

\author{Liang Jiang}

\affiliation{Departments of Applied Physics and Physics, Yale University, New
Haven, Connecticut 06520, USA}
\begin{abstract}
We investigate quantum control of an oscillator mode off-resonantly
coupled to an ancillary qubit. In the strong dispersive regime, we
may drive the qubit conditioned on number states of the oscillator,
which together with displacement operations can achieve universal
control of the oscillator. Based on our proof of universal control,
we provide explicit constructions for arbitrary state preparation
and arbitrary unitary operation of the oscillator. Moreover, we present
an efficient procedure to prepare the number state $\left|n\right\rangle $
using only $O\left(\sqrt{n}\right)$ operations. We also compare our
scheme with known quantum control protocols for coupled qubit-oscillator
systems. This universal control scheme of the oscillator can readily
be implemented using superconducting circuits.
\end{abstract}

\date{\today}


\maketitle
As an important model for quantum information processing, the coupled
qubit-oscillator system has been actively investigated in various
platforms, including trapped ions \cite{Leibfried03RMP}, nano-photonics
\cite{Tiecke14}, cavity QED \cite{Reiserer14}, and circuit QED \cite{Devoret13}.
Due to its convenient control, the physical qubit is usually the primary
resource for quantum information processing. Meanwhile, the oscillator
serves as an auxiliary system for quantum state transfer and detection
\cite{Schoelkopf08}. In fact, the oscillator, associated with the
phononic or photonic mode, may have long coherent times \cite{Leibfried03RMP,Palomaki13,Reagor13}
and the large Hilbert space associated with the oscillator can be
used for quantum encoding \cite{Gottesman01a,Leghtas13a,Leghtas13b}
and autonomous error correction with engineered dissipation \cite{Mirrahimi14}.
These crucial features call for deeper investigations into quantum
control theory of an oscillator.

The seminal work by Law and Eberly \cite{Law96} has triggered many
theoretical and experimental investigations to prepare quantum states
of the oscillator assisted by an ancillary qubit with Jaynes-Cummings
(JC) coupling \cite{Brattke01,Leibfried03RMP,Houck07,Hofheinz09},
while the general problem of implementing arbitrary unitary operation
remains an outstanding challenge. Even with recent advances, protocols
for universal control require either a large number of control operations
\cite{Mischuck13} or a more complicated model with an ancillary three-level
system \cite{Santos05}.  Meanwhile, development in superconducting
circuits acting in the strong dispersive regime opens new possibilities
for universal control of the oscillator \cite{Schuster07}.

In this Letter, we provide schemes for arbitrary state preparation
and universal control of the oscillator assisted by an ancillary qubit.
These schemes utilize the dispersive Hamiltonian \cite{Schuster07}
along with two types of drives associated with the qubit and the oscillator,
respectively. The key is the capability to drive the qubit \cite{Schuster07,Johnson10,Leghtas13a,Leghtas13b,Vlastakis13,NiggS14}
and impart arbitrary phases conditioned on the number state of the
oscillator. 

\begin{figure}
\begin{centering}
\includegraphics[width=8.5cm]{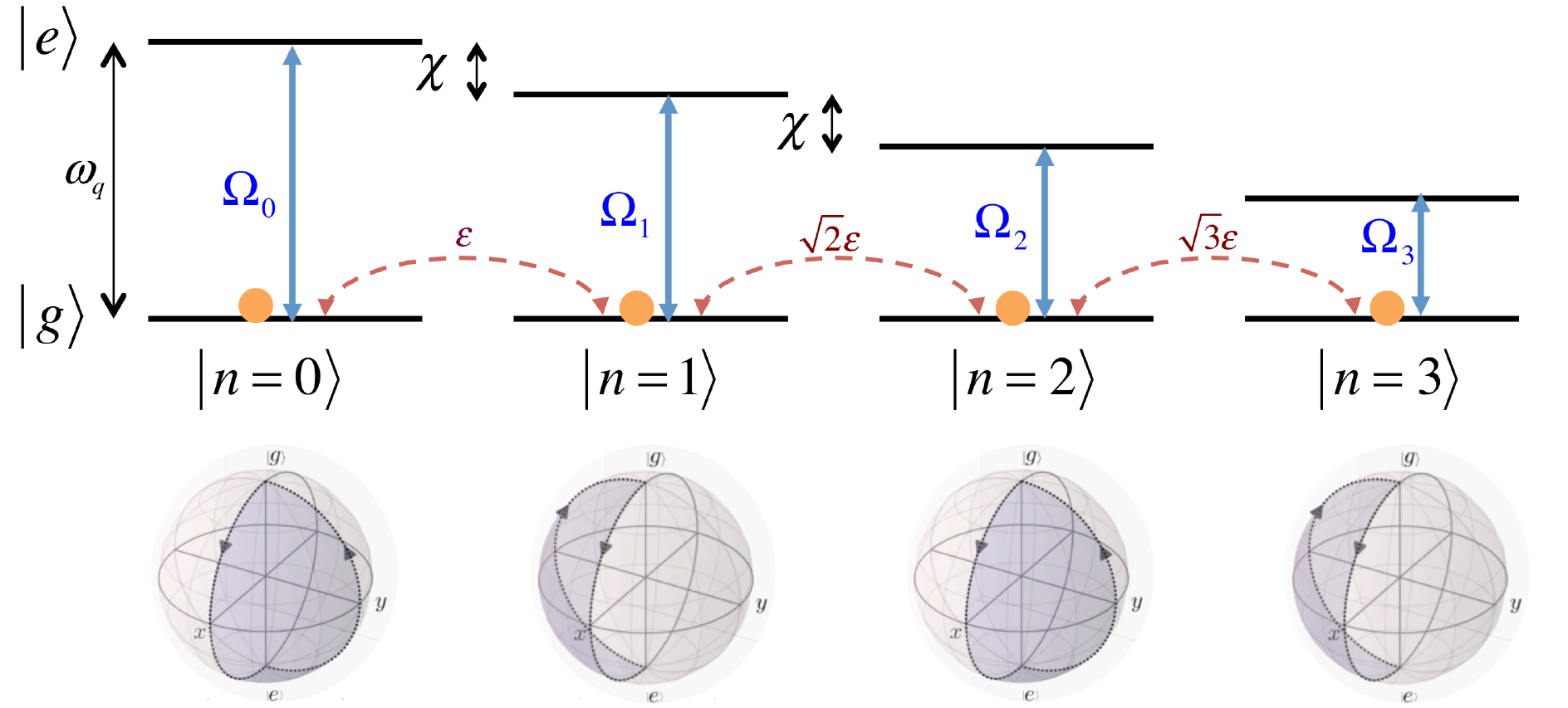}
\par\end{centering}

\protect\caption{\label{fig:operations}Energy level diagram of qubit-oscillator system.
In the rotating frame of the oscillator, the states $\left\{ \left|g,n\right\rangle \right\} _{n}$
have the same energy. After each operation, the population (orange
circles) remains in the subspace associated with $\protect\ket g$.
The displacement operation (red dashed arrows) couples the states
$\left|g,n-1\right\rangle $ and $\left|g,n\right\rangle $ with strength
$\sqrt{n}\varepsilon$ for all $n$. The SNAP gate (blue solid arrows)
can simultaneously accumulate different Berry phases $\left\{ \theta_{n}\right\} $
to states $\left\{ \left|g,n\right\rangle \right\} $. The Berry phase
$\theta_{n}$ is proportional to the enclosed shaded area in the corresponding
Bloch sphere, achieved by resonant microwave pulses with frequency
$\omega_{q}-n\chi$ (blue oscillatory fields). }
\end{figure}

The Hamiltonian of the qubit-oscillator system is 
\begin{eqnarray}
\hat{H} & = & \hat{H}_{0}+\hat{H}_{1}+\hat{H}_{2},\label{eq:bare_hamiltonian}
\end{eqnarray}
with dispersively coupled qubit and oscillator \cite{Schuster07}
\begin{equation}
\hat{H}_{0}=\omega_{q}\mid e\rangle\langle e\mid+\omega_{c}\ph-\chi\mid e\rangle\langle e\mid\ph,\label{eq:H0}
\end{equation}
time-dependent drive of the oscillator
\begin{equation}
\hat{H}_{1}=\epsilon\left(t\right)e^{i\omega_{c}t}\aa^{\dagger}+h.c.,\label{eq:H1}
\end{equation}
and time-dependent drive of the qubit
\begin{eqnarray}
\hat{H}_{2} & = & \Omega\left(t\right)e^{i\omega_{q}t}\left|e\rangle\langle g\right|+h.c.\label{eq:H2}
\end{eqnarray}
where $\omega_{q}$ is the qubit transition frequency between $\ket g$
and $\ket e$, $\omega_{c}$ is the oscillator frequency, $a^{\dagger}$($a$)
are the raising (lowering) operators, $\hat{n}=\hat{a}^{\dagger}\hat{a}$
is the number operator of the oscillator, $\chi$ is the dispersive
coupling, $\Omega(t)$ and $\epsilon(t)$ are the time-dependent drives
of the qubit and the oscillator, respectively. The eigenstates of
$\hat{H}_{0}$ are $\ket{g,n}$ and $\ket{e,n}$ with oscillator excitation
number $n=0,1,\cdots$, as illustrated in Fig.~\ref{fig:operations}.

We consider control schemes with three constraints to achieve universal
control of the oscillator:
\begin{enumerate}
\item The oscillator and qubit are never driven simultaneously (i.e. $\epsilon(t)\Omega(t)=0$
for all $t$);
\item The qubit is in the ground state $\left|g\right\rangle $ whenever
the oscillator drive is on (i.e. when $\epsilon(t)\ne0$);
\item The qubit drive is weak compared with the dispersive coupling, i.e.
$\left|\Omega(t)\right|\ll\chi$.
\end{enumerate}
With the above constraints, we have two types of operations. The type
$\one$ operation (based on Eq.~(\ref{eq:H1}) under constraint \#2)
is a displacement operation 
\begin{equation}
\one\,\,\hat{D}(\alpha)\,\,=\exp\left(\alpha\aa^{\dagger}-\alpha^{*}\aa\right),\label{eq:Displacement}
\end{equation}
with $\alpha=i\int\varepsilon(t)dt$, which can coherently pump or
remove energy from the oscillator. The type $\two$ operation (based
on Eq.~(\ref{eq:H2}) under constraint \#3) is a qubit rotation \textit{conditional
on the number states }\cite{Schuster07,Johnson10,Vlastakis13,NiggS14,NiggS13},
which can impart number dependent Berry phases. As illustrated in
Fig.~\ref{fig:operations}, there is a series of transition frequencies
of the qubit, $\left\{ \omega_{q}-\chi n\right\} _{n=0}^{\infty}$,
depending on the excitation number of the oscillator. We can achieve
unitary rotations between the selected levels $\left\{ \left|g,n\right\rangle ,\left|e,n\right\rangle \right\} $
with negligible effect to the rest of system, if we drive the qubit
with $\Omega\left(t\right)=\Omega_{n}\left(t\right)e^{-in\chi t}$
and $\left|\Omega_{n}\right|\ll\chi$ \cite{Schuster07,Johnson10,Vlastakis13,NiggS14}.
Hence, we can impart a Berry phase to a selected number state: $\left|g,n\right\rangle \rightarrow e^{i\theta_{n}}\left|g,n\right\rangle $,
with $\theta_{n}$ proportional to the solid angle subtended by the
path in the Bloch sphere associated with $\left\{ \left|g,n\right\rangle ,\left|e,n\right\rangle \right\} $
(as illustrated in Fig. \ref{fig:operations}). Since the qubit remains
in $\ket g$ after the operation, we can effectively obtain a \emph{Selective
Number-dependent Arbitrary Phase} (SNAP) operation
\begin{equation}
\hat{S}_{n}\left(\theta_{n}\right)=e^{i\theta_{n}\left|n\rangle\langle n\right|},
\end{equation}
which imparts phase $\theta_{n}$ to the number state $\left|n\right\rangle $.
Since the excitation number is preserved during the SNAP operation
(due to constraint \#1), we may drive the qubit with multiple frequency
components, $\Omega\left(t\right)=\sum_{n}\Omega_{n}\left(t\right)e^{i\left(\omega_{q}-\chi n\right)t}$.
These will simultaneously accumulate different phases $\theta_{n}$
for different number states and implement the general SNAP gate 
\begin{equation}
\two\,\,\hat{S}\left(\vec{\theta}\right)\,=\prod_{n=0}^{\infty}\hat{S}_{n}\left(\theta_{n}\right)\,=\sum_{n=0}^{\infty}e^{i\theta_{n}}\mid n\rangle\langle n\mid,\label{eq:SNAP}
\end{equation}
where $\vec{\theta}=\left\{ \theta_{n}\right\} _{n=0}^{\infty}$ is
the list of phases. Since $\theta_{n}$ can be an arbitrary function
of $n$, the SNAP gate can simulate arbitrary non-linear effects that
conserve the excitation number. For example, if we choose $\theta_{n}\propto n^{2}$,
the SNAP gate effectively induces a Kerr nonlinearity of the oscillator.
With SNAP gates, we just need to consider real displacement, because
any complex displacement $\alpha=re^{i\phi}$ can be decomposed as
real displacement and SNAP gates, $\hat{D}(\alpha)=\hat{S}(\vec{\theta})\hat{D}(r)\hat{S}(-\vec{\theta}),$
with $\theta_{n}=n\phi\left(\mbox{mod }2\pi\right)$.

\paragraph{Proof of universality.}

To show that the operations $\hat{D}(\alpha)$ and $\hat{S}(\vec{\theta})$
are sufficient for universal control of the oscillator, we first
identify $\hat{p}=-i(\aa^{\dagger}-\aa)$ as a generator of $\hat{D}\left(\alpha\right)$
for real $\alpha$, and $\left\{ \hat{Q}_{n}=\sum_{n'=0}^{n}\left|n'\rangle\langle n'\right|\right\} _{n}$
as generators of $\hat{S}(\vec{\theta})$. Their commutator is 
\[
\hat{J}_{n}=i\left[\hat{p},\hat{Q}_{n}\right]=\sqrt{n+1}\left(\mid n\rangle\langle n+1\mid+\mid n+1\rangle\langle n\mid\right),
\]
which can selectively couple $n$ and $n-1$. This gives the group
commutator 
\begin{eqnarray}
 &  & {\textstyle \hat{D}(\epsilon)\hat{R}_{n}\left(\epsilon\right)\hat{D}(-\epsilon)\hat{R}_{n}\left(-\epsilon\right)}=\exp\left(iJ_{n}\epsilon^{2}+\mathcal{O}(\epsilon^{3})\right),\label{eq:socks_and_shoes}
\end{eqnarray}
for small real $\epsilon$ and for the SNAP gate $\hat{S}\left(\{\epsilon,\dots\epsilon,0,\dots\}\right)=\hat{R}_{n}\left(\epsilon\right)=e^{iQ_{n}\epsilon}=\sum_{n'=0}^{n}e^{i\epsilon}\left|n'\rangle\langle n'\right|+\sum_{n'=n+1}^{\infty}\left|n'\rangle\langle n'\right|$.
For any integer $N>0$, $\left\{ \hat{J}_{n}\right\} _{n=0}^{N-1}$
and $\left\{ \hat{Q}_{n}\right\} _{n=0}^{N-1}$ are sufficient to
generate the Lie algebra $\mathfrak{u}(N)$ over the truncated number
space spanned by $\left\{ \ket n|n<N\right\} $, which implies universal
control of the oscillator \cite{Lloyd99,Braunstein05,jacobs2007engineering}.

\begin{table}
{\tiny{}}\subfloat[\label{tab:n_to_n+1}]{\begin{centering}
{\footnotesize{}$\qquad$}%
\begin{tabular}{|c|c|c|c|c|}
\hline 
{\footnotesize{}$n$} & {\footnotesize{}$\alpha_{n,1}$} & {\footnotesize{}$\alpha_{n,2}$} & {\footnotesize{}$\alpha_{n,3}$} & {\footnotesize{}$1-F$}\tabularnewline
\hline 
\hline 
{\footnotesize{}0} & {\footnotesize{}-0.575} & {\footnotesize{}0.682} & {\footnotesize{}-0.371} & {\footnotesize{}$8.3\times10^{-4}$}\tabularnewline
\hline 
{\footnotesize{}1} & {\footnotesize{}-0.313} & {\footnotesize{}0.539} & {\footnotesize{}-0.316} & {\footnotesize{}$6.4\times10^{-4}$}\tabularnewline
\hline 
{\footnotesize{}2} & {\footnotesize{}-0.256} & {\footnotesize{}0.441} & {\footnotesize{}-0.258} & {\footnotesize{}$5.2\times10^{-4}$}\tabularnewline
\hline 
{\footnotesize{}3} & {\footnotesize{}-0.222} & {\footnotesize{}0.382} & {\footnotesize{}-0.223} & {\footnotesize{}$4.9\times10^{-4}$}\tabularnewline
\hline 
{\footnotesize{}4} & {\footnotesize{}-0.198} & {\footnotesize{}0.341} & {\footnotesize{}-0.200} & {\footnotesize{}$4.7\times10^{-4}$}\tabularnewline
\hline 
{\footnotesize{}5} & {\footnotesize{}-0.181} & {\footnotesize{}0.312} & {\footnotesize{}-0.182} & {\footnotesize{}$4.6\times10^{-4}$}\tabularnewline
\hline 
\end{tabular}
\par\end{centering}{\footnotesize \par}

{\tiny{}}}{\tiny{}}\subfloat[\label{tab:0_to_sum}]{\centering{}{\footnotesize{}$\qquad$}%
\begin{tabular}{|c|c|}
\hline 
{\footnotesize{}$N$} & {\footnotesize{}$1-F$}\tabularnewline
\hline 
\hline 
{\footnotesize{}1} & {\footnotesize{}$2.7\times10^{-5}$}\tabularnewline
\hline 
{\footnotesize{}2} & {\footnotesize{}$1.8\times10^{-5}$}\tabularnewline
\hline 
{\footnotesize{}3} & {\footnotesize{}$1.1\times10^{-5}$}\tabularnewline
\hline 
{\footnotesize{}4} & {\footnotesize{}$7.2\times10^{-6}$}\tabularnewline
\hline 
{\footnotesize{}5} & {\footnotesize{}$5.2\times10^{-6}$}\tabularnewline
\hline 
{\footnotesize{}6} & {\footnotesize{}$4.0\times10^{-6}$}\tabularnewline
\hline 
\end{tabular}{\footnotesize{}$\qquad$}{\tiny{}}}\protect\caption{\textbf{(a)} Fidelity and optimized displacements associated with
state preparation from $\left|\psi_{\mathrm{init}}\right\rangle =\left|n\right\rangle $
to $\left|\psi_{\mathrm{final}}\right\rangle =\left|n+1\right\rangle $.
\textbf{(b)} Fidelity associated with state preparation from $\left|\psi_{\mathrm{init}}\right\rangle =\left|0\right\rangle $
to $\left|\psi_{\mathrm{final}}\right\rangle =\frac{1}{\sqrt{N+1}}\sum_{n=0}^{N}\left|n\right\rangle $.}
\end{table}

\paragraph{Explicit construction of target state.}

Instead of using infinitesimal evolutions suggested by the universality
proof, we would like to construct a control sequence with finite number
of steps to prepare the target state $\left|\psi\right\rangle =\sum_{n=0}^{N}c_{n}\left|n\right\rangle $
from the oscillator ground state $\left|0\right\rangle $. First,
we observe that the target state can be expressed as $\left|\psi\right\rangle =\tilde{S}|\tilde{\psi}\rangle$,
where $\tilde{S}=\hat{S}\left(\left\{ \text{angle}\left(c_{n}\right)\right\} _{n}\right)$
is the SNAP gate and $|\tilde{\psi}\rangle=\sum_{n=0}^{N}\tilde{c}_{n}\left|n\right\rangle $
with $\tilde{c}_{n}=\left|c_{n}\right|$. Hence, we only need to consider
the new target state $|\tilde{\psi}\rangle$ with non-negative $\tilde{c}_{n}$
in the number basis. We then take the strategy to ``unroll'' the
amplitude via a sequence of intermediate states $\left\{ |\tilde{\psi}_{n}\rangle\right\} _{n=0}^{N}$,
with $|\tilde{\psi}_{n}\rangle=\left(\sum_{n'=0}^{n-1}\tilde{c}_{n'}\ket{n'}\right)+\tilde{d}_{n}\ket n$
and $\tilde{d}_{n}=\sqrt{\sum_{n'=n}^{N}\tilde{c}_{n'}^{2}}$, which
connects the initial state $|\tilde{\psi}_{0}\rangle=\ket 0$ and
the target state $|\tilde{\psi}_{N}\rangle=|\tilde{\psi}\rangle$.
The key is to perform a rotation $\hat{U}_{n}\in SO(2)$ that acts
non-trivially in the subspace spanned by $\left\{ \left|n\right\rangle ,\left|n+1\right\rangle \right\} $,
so that $\hat{U}_{n}\tilde{d}_{n}\ket n=\tilde{c}_{n}\ket n+\tilde{d}_{n+1}\ket{n+1}$
and consequently $\hat{U}_{n}|\tilde{\psi}_{n}\rangle\approx|\tilde{\psi}_{n+1}\rangle$
for $n=0,1,\cdots,N$. 

Let us first consider the rotation $\hat{U}_{n}$ that transfers population
from $\left|\psi_{\mathrm{init}}\right\rangle =\left|n\right\rangle $
to $\left|\psi_{\mathrm{final}}\right\rangle =\left|n+1\right\rangle $,
with an efficient implementation 
\begin{equation}
\hat{U}_{n}=\hat{D}(\alpha_{1}^{\left(n\right)})\hat{R}_{n}\left(\pi\right)\hat{D}(\alpha_{2}^{\left(n\right)})\hat{R}_{n}\left(\pi\right)\hat{D}(\alpha_{3}^{\left(n\right)}),\label{eq:DRDRD}
\end{equation}
where $\hat{R}_{n}\left(\pi\right)=-\sum_{n'=0}^{n}\left|n'\rangle\langle n'\right|+\sum_{n'=n+1}^{\infty}\left|n'\rangle\langle n'\right|$
is a SNAP gate with $\pi$ phase shift for number states with no more
than $n$ excitations. To maximize the state preparation fidelity,
$F=\left|\bra{\psi_{\mathrm{\mathrm{final}}}}\hat{U}_{n}\ket{\psi_{\mathrm{\mathrm{init}}}}\right|$
, we obtain optimized displacements $\left(\alpha_{1},\alpha_{2},\alpha_{3}\right)$
as listed in Table \ref{tab:n_to_n+1}. Moreover, we can optimize
$\hat{U}_{n}$ for coherently transfer from $\left|\psi_{\mathrm{init}}\right\rangle =\left|n\right\rangle $
to $\left|\psi_{\mathrm{final}}\right\rangle =\sin\left(\theta\right)\left|n+1\right\rangle +\cos\left(\theta\right)\left|n\right\rangle $
for $\theta\in\left[0,\frac{\pi}{2}\right]$ with fidelity better
than $0.999$.

Using these building blocks we can first construct a sequence of operators
$\hat{U}_{n}\in SO(2)$ acting on the subspaces $\left\{ \left|n\right\rangle ,\left|n+1\right\rangle \right\} $
so that $\hat{U}_{n}\left|n\right\rangle =\cos\theta_{n}\left|n\right\rangle +\sin\theta_{n}\left|n+1\right\rangle $
with $\sin\theta_{n}=\frac{d_{n+1}}{d_{n}}$. The product $\hat{U}_{N-1}\cdots\hat{U}_{1}\hat{U}_{0}$
provides an initial guess for state preparation. We then combine the
displacement $\hat{D}(\alpha_{1}^{\left(n\right)})$ from $\hat{U}_{n}$
with the displacement $\hat{D}(\alpha_{3}^{\left(n+1\right)})$ from
$\hat{U}_{n+1}$ to reduce the number of parameters. Finally, we optimize
over all $2N+1$ displacement parameters with an initial guess based
on the result from the previous ``local'' optimizations \cite{Khaneja05}.
Performing the complete procedure requires $N$ optimizations over
3 parameters and one optimization over $2N+1$ parameters, with optimized
fidelity $F>0.999$ as listed in Table~\ref{tab:0_to_sum} for state
preparation from $\left|\psi_{\mathrm{init}}\right\rangle =\left|0\right\rangle $
to $\left|\psi_{\mathrm{final}}\right\rangle =\frac{1}{\sqrt{N+1}}\sum_{n=0}^{N}\left|n\right\rangle $.

\paragraph{Sublinear scheme to prepare number state.}

\begin{figure}
\centering{}\includegraphics[width=8cm]{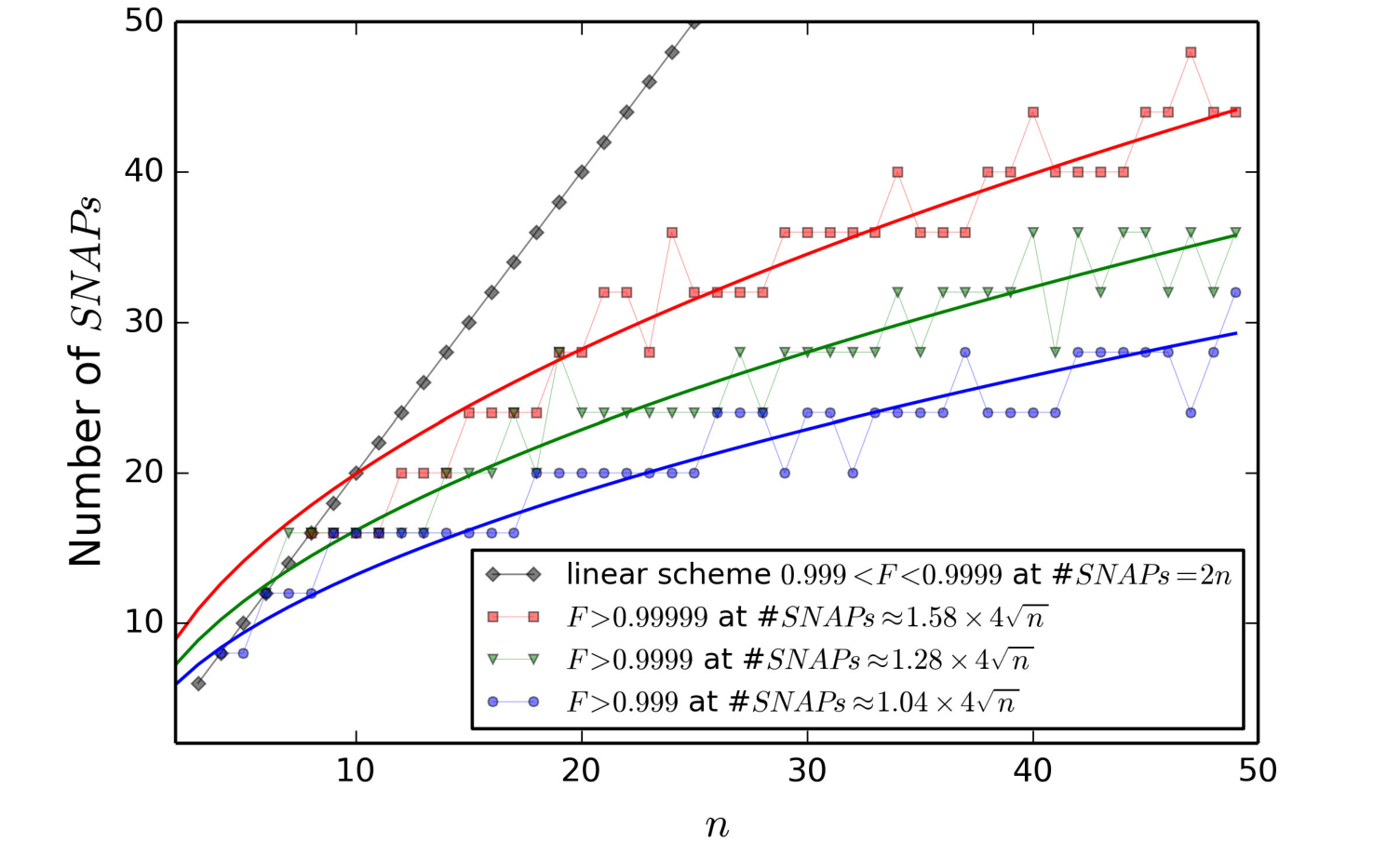}\protect\caption{The number of SNAP gates for the preparation from $\protect\ket 0$
to $\left|n\right\rangle $ (with fixed lower bound of fidelity) for
generic linear scheme (black diamond line) and specialized sublinear
schemes with different target fidelities (colored lines). The specialized
schemes operate on $D(\sqrt{n})\left|0\right\rangle $ by folding
all population from the subspace $\left\{ \left|n-\Delta n\right\rangle ,\dots,\left|n+\Delta n\right\rangle \right\} $
to the number state $\protect\ket n$. \label{fig:coh_to_n}}
\end{figure}

The above scheme of employing $SO(2)$ rotations to unroll the amplitudes
works generically for arbitrary target states, with the number of
gates scaling linearly with the highest excitation number. However,
certain states with a narrow distribution of photon numbers can be
prepared more efficiently by taking advantage of the fast experimentally-available
displacement operations. For example, the preparation of the number
state $\left|n\right\rangle $ requires $O\left(n\right)$ sequential
$SO\left(2\right)$ rotations from $\ket 0$ using the generic scheme.
In contrast, if we start from the coherent state $D(\alpha)\left|0\right\rangle =\left|\alpha\right\rangle $
with $\alpha=\sqrt{n}$, which has population distribution centered
around $\left|n\right\rangle $ with a spread of $\mathcal{O}\left(\sqrt{n}\right)$,
we need only $\mathcal{O}\left(\sqrt{n}\right)$ rounds of $SO(2)$
rotations to ``fold'' the coherent state $\ket{\alpha}$ to the
number state $\left|n\right\rangle $. Fig.~\ref{fig:coh_to_n} compares
the number of SNAP gates needed between the generic linear scheme
(with $O\left(n\right)$ operations) and the specialized sublinear
schemes (with $O\left(\sqrt{n}\right)$ operations) designed for preparation
from $\ket 0$ to $\ket n$, with various target fidelities. For $n\gtrsim8$,
it becomes advantageous to use the specialized sublinear scheme instead
of the generic scheme.

\paragraph{Explicit construction of target unitary.}

A more general version of the problem of arbitrary state preparation
is the construction of an arbitrary unitary operation of the oscillator.
The first step is to construct the $SO(2)$ transformation in the
subspace spanned by $\{\left|n\right\rangle ,\left|n+1\right\rangle \}$,
based on the similar construction as Eq.~(\ref{eq:DRDRD}). Different
from state preparation discussed earlier, here $\hat{U}_{n}$ has
to fulfill both conditions: $\hat{U}_{n}\left|n\right\rangle =\cos\theta_{n}\left|n\right\rangle +\sin\theta_{n}\left|n+1\right\rangle $
and $\hat{U}_{n}\left|n+1\right\rangle =-\sin\theta_{n}\left|n\right\rangle +\cos\theta_{n}\left|n+1\right\rangle $
with $SO(2)$ rotation angle $\theta_{n}$. In addition, it is important
to impose the constraint $\alpha_{1}+\alpha_{2}+\alpha_{3}=0$ to
Eq.~(\ref{eq:DRDRD}) to minimize the undesired effects to the subspace
associated with $\ket{n'}\neq\ket n$ or $\ket{n+1}$. Numerically,
we optimize the unitary operation fidelity, $F_{\mathrm{unitary}}=\frac{1}{N_{c}}\left|Tr\left(\hat{U}_{n}^{\dagger}\hat{U}_{n,\mathrm{ideal}}\right)\right|$,
where $N_{c}$ is the cutoff dimension (i.e. the size of the matrices
used to represent the operators) \cite{Khaneja05}%
\footnote{This definition of fidelity is slightly faster to calculate than the
definition with uniform norm, but more importantly it permits analytic
expressions for the gradient of the cost function, greatly speeding
up the optimization algorithm.%
}. The numerical optimum is attained with $\left(\alpha_{1},\alpha_{2},\alpha_{3}\right)=\left(\alpha,-2\alpha,\alpha\right)$.

\begin{figure}
\begin{centering}
\includegraphics[width=8cm]{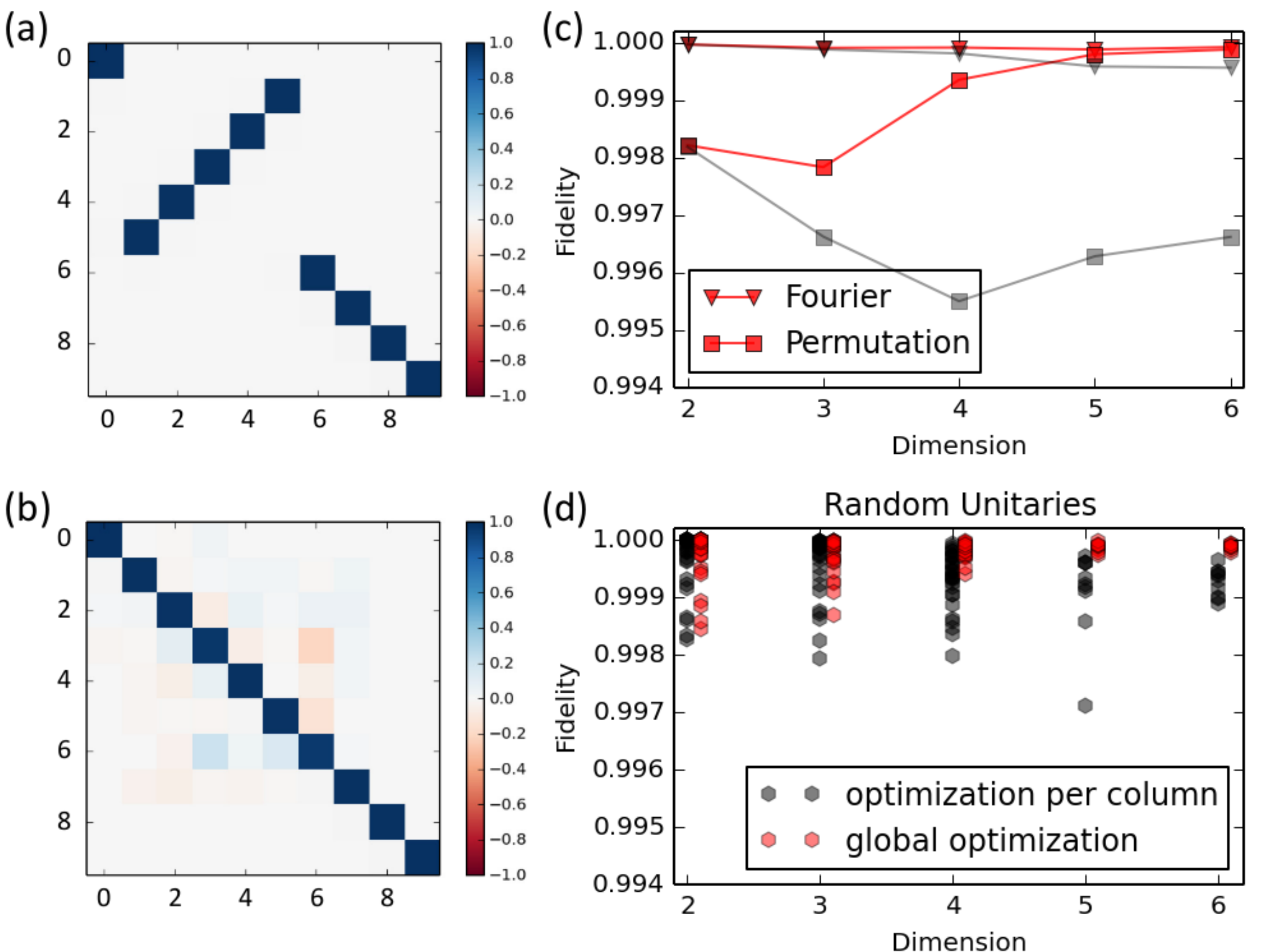}
\par\end{centering}

\protect\caption{\label{fig:un_example}\textbf{(a)} An example target unitary operation,
$U_{\mathrm{target}}$ (a permutation). \textbf{(b)} The product is
close to identity, $\hat{U}_{\mathrm{construct}}\hat{U}_{\mathrm{target}}^{-1}\approx\hat{I}$,
with small non-zero off-diagonal elements. \textbf{(c,d)} Comparison
of fidelities after column-wise optimization (black) and fidelities
with additional optimization over all displacement parameters (red)
for \textbf{(c)} Fourier (triangles) and permutation (squares) operations
and \textbf{(d)} randomly generated target unitary operations (hexagons). }
\end{figure}

To construct an arbitrary unitary $\hat{U}_{\mathrm{target}}$ in
the $\{\left|0\right\rangle ,\dots,\left|n-1\right\rangle \}$ subspace,
we start by taking its inverse:
\[
\hat{U}_{\mathrm{target}}^{-1}=\begin{pmatrix}\hat{W}_{n} & \vline & 0\\
\hline 0 & \vline & \hat{I}_{N_{c}-n}
\end{pmatrix},
\]
where $\hat{W}_{n}$ is the non-trivial block and $\hat{I}_{n'}$
is the $n'\times n'$ identity matrix. We first apply a SNAP gate
such that the last column of the $\hat{W}_{n}$ block now contains
only non-negative coefficients. We then apply $n-1$ consecutive $SO(2)$
rotations eliminating the off-diagonal elements in the last column
of $\hat{W}_{n}$, such that the column becomes $\left(0,\cdots,0,1\right)_{n}^{\mathsf{T}}$.
Since all rows of a unitary matrix are orthonormal, the last row of
the $\hat{W}_{n}$ block must be transformed to $\left(0,\cdots,0,1\right)_{n}$
. Hence, the result is 
\[
\hat{V}_{n-1}^{\left(n\right)}\hat{V}_{n-2}^{\left(n\right)}\cdots\hat{V}_{0}^{\left(n\right)}\hat{S}^{\left(n\right)}\hat{U}_{\mathrm{target}}^{-1}=\begin{pmatrix}\hat{W}_{n-1} & \vline & 0 & \vline & 0\\
\hline 0 & \vline & 1 & \vline & 0\\
\hline 0 & \vline & 0 & \vline & I_{N_{c}-n}
\end{pmatrix},
\]
where $\hat{S}^{\left(n\right)}$ is a SNAP gate necessary for any
complex phases unobtainable with the $SO(2)$ operations. The $n-1$
$SO(2)$ rotations 
\[
\hat{V}_{k}^{\left(n\right)}=\hat{D}(\alpha_{k}^{\left(n\right)})\hat{R}_{k}\left(\pi\right)\hat{D}(-2\alpha_{k}^{\left(n\right)})\hat{R}_{k}\left(\pi\right)\hat{D}(\alpha_{k}^{\left(n\right)})
\]
can be individually optimized, before being chained together for a
second round of optimization over $n-1$ displacement parameters $\left\{ \alpha_{k}^{\left(n\right)}\right\} _{k=0}^{n-1}$.
The cost function to be minimized for the second round of optimization
is the sum of absolute values of off-diagonal terms (excluding $\hat{W}_{n-1}$
block). We iterate the procedure until we obtain 
\begin{equation}
\hat{U}_{\mathrm{construct}}\hat{U}_{\mathrm{target}}^{-1}\approx\hat{I},
\end{equation}
with $\hat{U}_{\mathrm{construct}}=\prod_{n'=1}^{n}\left(\hat{V}_{n'-1}^{\left(n'\right)}\hat{V}_{n'-2}^{\left(n'\right)}\cdots\hat{V}_{0}^{\left(n'\right)}\hat{S}^{\left(n'\right)}\right)$,
as illustrated in Fig.~\ref{fig:un_example}(a,b) for a specific
$U_{\text{target}}$ %
\footnote{The SciPy\cite{scipy} routines were used for numerical optimizations. %
} .

Using the above decomposition, we need $n(n-1)/2$ $SO(2)$ rotations
(each containing 3 displacements and 2 SNAP gates). We can combine
consecutive displacements, lowering the number of displacements to
2 per $SO(2)$ rotation. We also need one SNAP gate at each iteration
of the $\hat{W}_{n}\rightarrow\hat{W}_{n-1}$ step, with a total of
$n$ additional SNAP gates. For various $U_{\mathrm{target}}$, as
illustrated in Fig.~\ref{fig:un_example}(c,d), we find that the
step-wise optimization procedure can yield good final fidelity $F_{\mathrm{unitary}}>0.99$,
which can be further improved to $F_{\mathrm{unitary}}>0.999$ with
a third round of simultaneous optimization over all $n\left(n-1\right)/2$
displacement parameters $\{\alpha_{k}^{\left(n'\right)}\}_{k<n'\le n}$.

Our scheme can be applied to the general Hamiltonian with dispersive
coupling, $-\sum_{n}\chi_{n}\ket{e,n}\bra{e,n}$, as long as the number-dependent
qubit frequency shift can be resolved ($\left|\chi_{n}-\chi_{n'\neq n}\right|\gg\left|\Omega\left(t\right)\right|,\gamma,\kappa$,
for all relevant $n$ and $n'$, where $\gamma$ and $\kappa$ are
the qubit and cavity decoherence rates, respectively). Furthermore,
we can also extend the arbitrary unitary control to the subspaces
spanned by $\left\{ \left|g,n'\right\rangle ,\left|e,n'\right\rangle \right\} _{n'=0}^{n-1}$,
so that we can control the entire $2n$-dimensional quantum system.

\paragraph{Discussions.}

We now compare our SNAP-gate-based quantum control scheme with previous
protocols. The scheme proposed by Law and Eberly \cite{Law96} is
based on the JC model, $H_{\mathrm{JC}}\propto a\sigma^{+}+h.c.$,
which enables preparation of arbitrary superposition of number states.
The scheme by Mischuck and Molmer \cite{Mischuck13} further extended
the JC model from state preparation to arbitrary unitary operation,
but it is rather complicated because any JC control pulse necessarily
couples the states $\ket{g,n}$ and $\ket{e,n-1}$ (for all $n$)
simultaneously and with varying strength $g\sqrt{n}$. In contrast,
our scheme is based on the dispersive qubit-oscillator coupling, $H_{\mathrm{dispersive}}=-\chi\ket e\bra e\hat{n}$,
which preserves the oscillator number states, enables the SNAP gate
to directly access the two selected sublevels $\ket{g,n}$ and $\ket{e,n}$
with negligible effects to the rest of the levels, and ultimately
leads to efficient universal control of the oscillator. Similarly,
the proposal by Santos \cite{Santos05} introduces a different model
with a three-level $\Lambda$-type ancillary system to achieve universal
control, but it is experimentally more challenging than the simple
qubit ancilla considered in our scheme.

With dispersive qubit-oscillator coupling, there are other control
protocols available. For example, in the presence of the oscillator
drive $\epsilon$, we may ``block'' the processes $\left|n'\pm1\right\rangle \rightarrow\left|n'\right\rangle $
by driving the qubit resonantly to the transition $\ket{g,n'}\leftrightarrow\ket{e,n'}$,
with $\Omega_{n'}\gg\epsilon\sqrt{n'}$ \cite{Signoles14}. Similarly,
by resonantly driving transitions $\ket{g,n'}\leftrightarrow\ket{e,n'}$
for $n'<n$ and $n'>n+1$, we block all number changing transitions,
except for the transition between $\left\{ \left|n\right\rangle ,\left|n+1\right\rangle \right\} $
that can be used for $SO\left(2\right)$ unitary rotations. This blockade
scheme is relatively slow and each elementary operation takes time
$\tau\sim\left(\epsilon\sqrt{n}\right)^{-1}\gg\Omega^{-1}\gg\chi^{-1}$
due to the blockade requirements, while the SNAP-gate-based scheme
can be much faster with $\tau\sim\chi^{-1}$ by relaxing contraint
\#3 and numerically optimizing the shaped pulses with $\Omega\left(t\right)\sim\chi$.

In conclusion, the SNAP-gate-based scheme provides universal control
of the oscillator mode with strong dispersive coupling to a qubit.
Based on the proof of universal control, we show explicit constructions
for arbitrary state preparation and arbitrary unitary operation of
the oscillator. We also present an efficient procedure to prepare
the number state $\left|n\right\rangle $ using only $O\left(\sqrt{n}\right)$
operations. We note that deterministic SNAP-gate-based preparation
of $\left|n=1\right\rangle $ photon number state has been demonstrated
using superconducting circuits \cite{Heeres15}. The techniques introduced
here are not restricted to oscillator modes such as mechanical motions
\cite{Leibfried03RMP} and optical/microwave cavities \cite{Reiserer14,Tiecke14,Devoret13},
but can be extended to multi-level systems such as Rydberg atoms with
large angular momentum \cite{Signoles14}, as long as the dispersive
coupling between the qubit and oscillator/multi-level system is strong. 
\begin{acknowledgments}
We thank Michel H. Devoret, Luigi Frunzio, Steven M. Girvin, Zaki
Leghtas, Mazyar Mirrahimi, Andrei Petrenko, Matthew Reagor for helpful
discussions. The work was supported by ARO, AFOSR MURI, DARPA Quiness
program, NBRPC 973 program, the Alfred P. Sloan Foundation, and the
Packard Foundation. VVA acknowledges support from NSF GRFP (DGE-1122492).
We thank the Yale High Performance Computing Center for use of their
resources.
\end{acknowledgments}

\bibliographystyle{apsrev4-1}
\bibliography{library}

\end{document}